\documentclass[prl,twocolumn,superscriptaddress,showpacs,amscd,amsmath,amssymb,verbatim]{revtex4}

\usepackage{graphicx}
\usepackage[dvips]{hyperref}
\usepackage{textcomp}
\usepackage[OT1,T1]{fontenc}

\begin{document}
\newcommand{\kagome}{kagom\'e }

\title{Easy-Axis Kagom\'e Antiferromagnet: Local-Probe Study of Nd$_3$Ga$_5$SiO$_{14}$}
\author{A. Zorko}
\affiliation{Laboratoire de Physique des Solides,  Universit\'e Paris-Sud 11, UMR CNRS 8502, 91405
Orsay, France}
\affiliation{Jo\v{z}ef Stefan Institute, Jamova 39, 1000 Ljubljana, Slovenia}
\author{F. Bert}
\affiliation{Laboratoire de Physique des Solides,  Universit\'e Paris-Sud 11, UMR CNRS 8502, 91405
Orsay, France}
\author{P. Mendels}
\affiliation{Laboratoire de Physique des Solides,  Universit\'e Paris-Sud 11, UMR CNRS 8502, 91405
Orsay, France}
\author{P. Bordet}
\affiliation{Institut N\'eel, CNRS, BP 166, 38042 Grenoble, France}
\author{P. Lejay}
\affiliation{Institut N\'eel, CNRS, BP 166, 38042 Grenoble, France}
\author{J. Robert}
\affiliation{Institut N\'eel, CNRS, BP 166, 38042 Grenoble, France}

\date{\today}
\begin{abstract}
We report a local-probe investigation of the magnetically anisotropic kagom\'e compound Nd$_3$Ga$_5$SiO$_{14}$. Our zero-field $\mu$SR results provide a direct evidence of a fluctuating collective paramagnetic state down to 60~mK, supported by a wipe-out of the Ga nuclear magnetic resonance (NMR) signal below 25~K. At 60~mK a dynamics crossover to a much more static state is observed by $\mu$SR in magnetic fields above 0.5~T. Accordingly, the NMR signal is recovered at low $T$ above a threshold field, revealing a rapid temperature and field variation of the magnetic fluctuations.
\end{abstract}
\maketitle

Kagom\'e antiferromagnet (KAF), a 2D net of corner sharing triangles, is one of the most intensively studied examples of geometrically frustrated lattices. In the classical case, extensive degeneracy -- a hallmark of frustration -- is present in the ground state (GS) manifold, both in the Heisenberg and the Ising limit \cite{Chalker,Barry}. For the Heisenberg case, "flexible" coplanar configurations which allow for zero-energy weathervane modes, are likely selected by thermal fluctuations \cite{Chalker}. Frustration is even more severe in the Ising limit, where there are minimal spin degrees of freedom to accommodate the competing individual interactions. The system remains disordered at all temperatures, with exponentially decaying pair spin correlations \cite{Barry}. In the intermediate case, various anisotropy terms \cite{Kuroda, Elhajal, Sen} adding to the Heisenberg Hamiltonian can become important in selecting a particular GS subspace. Anisotropy and additional magnetic field $H$, favoring collinear spin configurations, may lead to particularly rich phase diagrams, as, e.g., predicted for the spin-1 KAF \cite{Xu}. In this context the exciting and mostly unexplored field of KAF with a sizable easy-axis anisotropy, could well yield novel original states.

\begin{figure}[b]
\includegraphics[width=8.2cm]{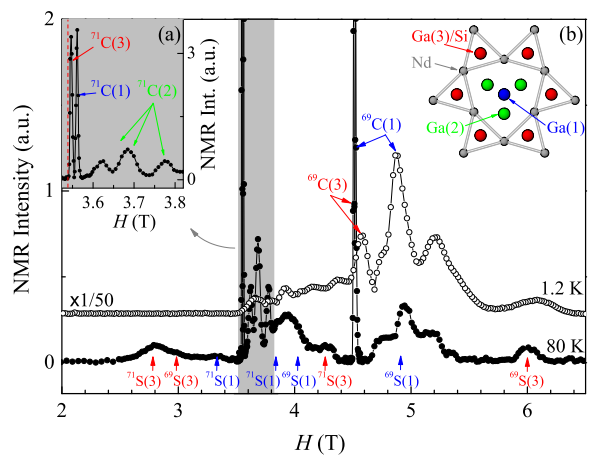}
\caption{(color online). $^{69, 71}$Ga NMR spectra of NGS at 46.01~MHz and $\textbf{H}||\textbf{c}$. The 1.2~K spectrum is translated vertically. Arrows indicate the satellite (S) lines at 80 K.  The inset (a) highlights central (C) peaks of the three crystallographically non-equivalent Ga sites at 80~K; the vertical dashed line corresponds to a reference field. The inset (b) shows the network of Nd$^{3+}$ magnetic moments and the location of the in-plane Ga(1) sites and Ga(2, 3) sites separating kagom\'e planes.}
\label{fig-1}
\end{figure}

\begin{figure}[t]
\includegraphics[width=8.0 cm]{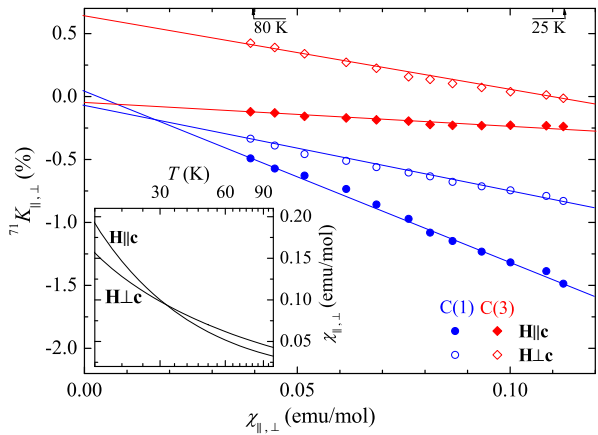}
\caption{(color online). Scaling of the $^{71}$Ga NMR shift with bulk susceptibility. Inset: The susceptibility crossover at 1~T.}
\label{fig-2}
\end{figure}

In the recently discovered Nd$_3$Ga$_5$SiO$_{14}$ (NGS) \cite{Bordet}, Nd$^{3+}$ ($S=9/2$) magnetic moments occupy a network of corner-sharing equilateral triangles, topologically equivalent to the perfect kagom\'e lattice if the nearest-neighbor exchange is considered [inset (b) to Fig.~\ref{fig-1}]. Susceptibility curves \cite{Bordet} display an interesting crossover from an easy-plane to an easy-axis ({\it c} axis, perpendicular to kagom\'e planes) behavior below 33~K, as shown in the inset to Fig.~\ref{fig-2}. Because of this anisotropy and large Nd$^{3+}$ moments, NGS is unique among kagom\'e-like compounds intensively studied in recent past \cite{Mendels}. The easy-axis low-$T$ behavior, originating from single-ion anisotropy and/or exchange anisotropy due to crystal-field (CF) effects, opens a route to experimental investigation of large-spin Ising-like KAF. The analysis of the high-$T$ susceptibility yielded a Curie-Weiss temperature $\theta_{CW}= -52~K$, although no irregularities were observed down to 1.6~K \cite{Bordet}. Interestingly, at this low $T$, a magnetization plateau was observed at around 1/2 of the saturated (free-ion) value in external fields $\textbf{H}||\textbf{c}$ of only few Tesla \cite{Bordet}.

In this Letter we report a series of complementary local-probe measurements, including nuclear magnetic resonance (NMR) and muon spin relaxation ($\mu$SR). Our results give a direct evidence of the fluctuating nature of the ground state and the absence of any static order in zero magnetic field. In addition, we investigate a dynamics crossover of the magnetic fluctuations when $\textbf{H}||\textbf{c}$ is applied, toward a more static polarized state. Our findings prove persistent spin fluctuations in the zero-field ground state down to 60~mK, characteristic of a spin-liquid state, which has been recently suggested from the absence of neutron magnetic Bragg peaks and the presence of a weak diffuse scattering down to 46~mK \cite{Zhou}. Partial ordering has been also reported under applied field at low $T$.

Polycrystalline Nd$_3$Ga$_5$SiO$_{14}$ samples were synthesized by a solid state reaction and a single crystal of cylindrical shape (2~mm in diameter and 8~mm in length) was grown by a floating zone technique \cite{Bordet}. NMR measurements were performed at various fixed frequencies by sweeping the magnetic field. Nuclear quadrupolar resonance (NQR) was also measured at zero field. Powder $\mu$SR measurements were performed on the LTF and GPS spectrometers at PSI, Switzerland.

We first report our Ga NMR investigation. The local magnetic coupling with the Nd$^{3+}$ moments is different for the three crystallographically non-equivalent gallium sites Ga(1,~2,~3) [inset (b) to Fig.~\ref{fig-1}]. In addition, both $^{69,71}$Ga isotopes contribute to the NMR spectrum. Through their sizable quadrupolar moments they are sensitive to local charge distributions, creating crystal electrostatic field gradients (EFG). We assigned all the observed peaks shown in Fig.~\ref{fig-1} -- three sets of central (C) and two satellite (S) lines for each isotope -- by performing frequency and angular dependent measurements. We took advantage of the EFG tensors symmetry \cite{Zorko}, known gyromagnetic ratios ($^{69}\gamma/^{71}\gamma = 0.7871$) and the ratio of the quadrupolar moments ($^{69}Q/^{71}Q=1.583$). Since the Ga(3) site is randomly occupied by Ga$^{3+}$ and Si$^{4+}$ \cite{Bordet}, a large distribution of quadrupolar frequencies is expected. This results in broad S lines and a broad NQR spectrum \cite{Zorko}. Surprisingly, the distribution of local environments does not lead to glassy-like behavior at low $T$.

In order to eliminate the quadrupolar broadening and get the best
accuracy on the local magnetic properties, the magnetic field was
applied perpendicular to the kagom\'e planes
($\textbf{H}||\textbf{c}$), along a local three-fold rotational axis
for the Ga(1,~3) sites. This results in narrow C(1,~3) lines, on
which we now focus. Their shift $K= (H_0 - H_c)/H_c=A\chi_l+ K_0$ of
a resonance field $H_c$ from a reference field $H_0$ is proportional
to the local susceptibility $\chi_l$. It has three contributions,
which need to be considered in the coupling constant
$A=A_{hf}+A_{dd}+A_{dm}$; a hyperfine ($hf$) and a dipolar ($dd$)
coupling between Nd$^{3+}$ moments and Ga nuclei, and a macroscopic
demagnetization ($dm$) field. $K_0$ is a $T$-independent chemical
shift. Our calculation of the dipolar field \cite{foot3} yields that
all these contributions are of a similar magnitude and that the
hyperfine coupling is almost isotropic \cite{Zorko}. In
Fig.~\ref{fig-2} we show the linear scaling of the shift with bulk
susceptibility between 80 K and 25 K. This proves that the local and
the bulk magnetic response are identical and, in particular, the
33~K crossing is also detected by our local-probe measurement. On
the contrary, in formerly studied 3$d$ metal kagom\'e-like compounds
the intrinsic susceptibility is regularly overshadowed by magnetic
defects contributions at low $T$ \cite{Mendels}.

\begin{figure}[b]
\includegraphics[width=8.4cm]{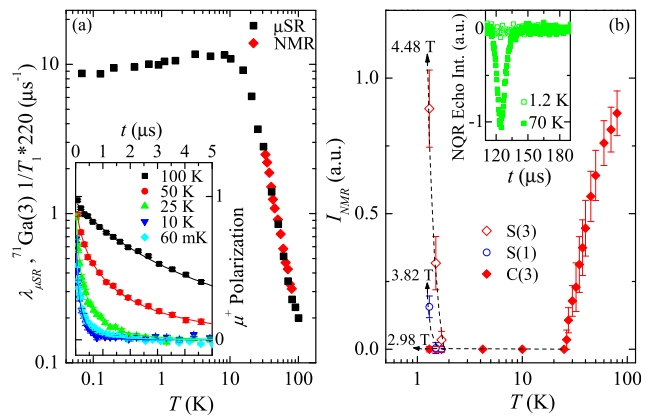}
\caption{(color online). (a) Zero-field (ZF) muon spin relaxation rate $\lambda$ and $^{71}$C(3) NMR spin-lattice relaxation rate $1/T_1$ measured at 46.01~MHz (3.55~T) and $\textbf{H}||\textbf{c}$. Inset: ZF relaxation of muon polarization. (b) Wipe-out of the NMR intensity. The high-$T$ (low-$T$) data were taken at 46.01~MHz (29.61~MHz) at different fields. Dashed lines are a guide to the eyes. Inset: Absence of the NQR echo at 1.2~K and 15~MHz.}
\label{fig-3}
\end{figure}

Next, we investigate the Nd$^{3+}$ spin dynamics. In Fig.~\ref{fig-3}(a) we show the $T$-dependence of the $^{71}$C(3) spin-lattice relaxation rate $1/T_1$ for $\textbf{H}||\textbf{c}$. Its values were obtained by fitting a magnetization recovery curve after saturation to the relaxation function $M_z(t)=M_z \left[ 1-0.4 {\rm exp}(-t/T_1) -0.6 {\rm exp}(-6t/T_1) \right]$, suited for magnetic relaxation of the central line in spin-3/2 nuclear systems \cite{Suter}. The relaxation is rather fast, proving that it is due to rapidly fluctuating dipolar and hyperfine fields, originating from Nd$^{3+}$ moments. With lowering $T$ the magnetic fluctuations are found to slow down considerably, which significantly enhances $1/T_1$ below 100~K and wipes out completely the signal below 25~K. The wipe-out is depicted in Fig \ref{fig-3}(b), displaying the $T$-dependence of the NMR intensity. It is caused by the increase of both the spin-lattice and the spin-spin relaxation rate. In a conventional transition scenario to a frozen state one would expect the NMR/NQR signal to emerge again below the transition temperature. In contrast, the inset to Fig.~\ref{fig-3}(b) demonstrates the absence of the NQR signal at 1.2~K, which points to persistent magnetic fluctuations. Similarly, the NMR signal is not detected at 1.2~K in small applied fields ($H\lesssim3$~T). Only in higher fields the NMR signal is recovered [Fig.~\ref{fig-1}, Fig.~\ref{fig-3}(b)]. We will address this field-induced effect later.

Further insight into the Nd$^{3+}$ dynamics is provided by
zero-field (ZF) $\mu$SR measurements. The muon relaxation rate
$\lambda$ could be followed in the full $T$-range from 100~K down to
60~mK and was extracted from a stretched-exponential fit $P(t)=P_0
{\rm exp}\left[ -(\lambda t)^\alpha \right]$ of the muon
polarization. The $T$-independent $\alpha=0.6(1)$ likely reflects a
variety of oxygen, hence $\mu^+$ sites in NGS. In the
simplified case of an exponential relaxation, $\lambda$ is related
to the Nd$^{3+}$ magnetic fluctuation rate $\nu_e$ by~\cite{Hayano}

\begin{equation}
\label{eq1} \lambda = \frac{
2\gamma_\mu^2H_\mu^2\nu_e}{\nu_e^2+\gamma_\mu^2 H_{LF}^2},
\end{equation}

\noindent where $H_\mu=0.2$~T is the average dipolar field created
by the Nd$^{3+}$ spins at the muon sites, $\gamma_\mu / 2 \pi=135.5$~MHz/T
is the muon gyromagnetic ratio and $H_{LF}$ stands for the applied
longitudinal field if any. The thermal evolution of $\lambda \propto
1/\nu_e$ in zero field, plotted in Fig.~\ref{fig-3}(a)~\cite{foot4},
reflects a strong slowing down of the Nd$^{3+}$ spin fluctuations -- from $\sim 200$~GHz at 100~K, in agreement with paramagnetic fluctuations governed by the magnitude of the exchange, to $\sim 4$~GHz at
10~K. The saturation of $\lambda$ below 10~K indicates that the slow
magnetic fluctuations, likely associated to strongly short-range
coupled Nd$^{3+}$ spins, persist in NGS down to 60~mK, which is a
fingerprint of a {\it dynamical ground state}, observed in many
kagom\'e-like \cite{Uemura,Keren,Keren2,Bono,Fukaya} and pyrochlore
\cite{Gardner,Dunsiger,Bert} lattices. In most of these materials
the dynamical plateau though coexists with a spin-glass-like
transition or even LRO. Such transitions are regularly observed at
the temperature of the plateau onset
\cite{Uemura,Keren,Keren2,Bono,Dunsiger,Bert}. Supported by a lack
of anomalies in the magnetic susceptibility of NGS \cite{Bordet},
the muon relaxation curves, which relax to zero down to 60~mK [inset
to Fig.~\ref{fig-3}(a)], provide a direct evidence of the {\it
absence of any static magnetic fields} emerging from the Nd$^{3+}$
moments. Our $\mu$SR data thus support {\it a collective paramagnetic
state} in NGS, stabilized below 10~K, which makes it together with the recently discovered spin-1/2
Heisenberg system ZnCu$_3$(OH)$_6$Cl$_2$ \cite{Shores} the only KAF
potentially displaying a pure spin-liquid ground state
\cite{Mendels2}.

\begin{figure}[t]
\includegraphics[width=8 cm]{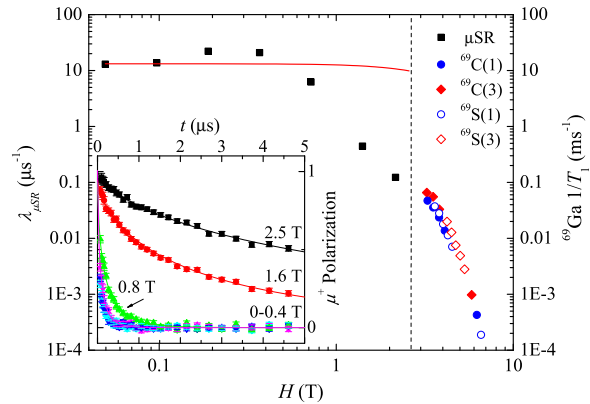}
\caption{(color online). Field dependence of the muon relaxation rate $\lambda$ at 60~mK and the $^{69}$Ga NMR $1/T_1$ at 1.2~K on several C and S lines. The solid line gives the predicted $H$-dependence of $\lambda$ in the dynamical case [Eq.~(\ref{eq1})]. Inset: Relaxation of muon polarization in various LF fields at 60~mK.}
\label{fig-4}
\end{figure}

When applying the longitudinal field (LF) at 60~mK, $\lambda$ varies moderately up to
0.5~T and then suddenly decreases by more than two orders of
magnitude for $H_{LF}=2.5$~T, as
visible already in raw relaxation curves (Fig. \ref{fig-4}). Such a strong response to $H_{LF}$, not observed
in any other kagom\'e-like compound \cite{Keren, Keren2, Fukaya},
cannot be explained within a fast fluctuation regime.
Eq.~(\ref{eq1}) predicts a smooth and much slower decrease of $\lambda$ with the upward shift of the Larmor frequency $\omega_L=\gamma_\mu H_{LF}$ with respect to the fixed fluctuation rate $\nu_e$
[solid line in Fig.~\ref{fig-4}]. The observed decrease of
$\lambda$ above $H_{LF}>0.5$~T demands that the spin dynamics
itself, $\nu_e$, is strongly reduced by $H_{LF}$, which suppresses the spectral density of fluctuations at $\omega_L$. Assuming
$H_\mu=0.2$~T, Eq.~(\ref{eq1}) gives $\nu_e\approx 10$~MHz at 2.5~T.
Therefore, our data point to a {\it field-induced crossover} from a
dynamical to a more static state for $H_{LF}>0.5$~T at 60~mK, in agreement with the recent detection of a static, net-ferromagnetic, state in the same $T$ and $H$ range~\cite{Zhou}.

The field-suppressed magnetic fluctuations are further reflected in
a pronounced $H$-dependence of the NMR spin-lattice relaxation rate
$1/T_1$ at 1.2~K, as measured at various fields in the spectra and
at various frequencies (Fig.~\ref{fig-4}). $1/T_1$ is lowered by
almost three orders of magnitude in a narrow $H$-window between 3~T
and 6.5~T. This enhanced effect of $H$ on the magnetic fluctuations
is responsible for only partial observability of the NMR spectrum at
low $T$. The 3.5~T threshold field at 1.2~K and 46.01~MHz is easily
extracted from the missing S(1,~3) lines below this field, as
evidenced in Fig.~\ref{fig-1}. The observed peaks are significantly
shifted and broadened with respect to the high-$T$ spectrum. Using
the shift of the narrowest $^{71}$C(1) line and its high-$T$
coupling constant $A$, the Nd$^{3+}$ magnetic moment of 1.6~$\mu_B$,
ca. half a value of the full moment, is extracted. Note that this is
the mean value of the six moments on the kagom\'e star, to which the
Ga(1) is coupled. Individual Nd$^{3+}$ moment could be larger than
1.6~$\mu_B$ in a non-ferromagnetic arrangement on the star. This
value is constant above the threshold field, in accordance with
previous magnetization measurements \cite{Bordet}. The static
line-shift and the dynamical relaxation thus exhibit strikingly
different $H$-dependence at 1.2~K. When lowering the temperature
below the wipe-out the NMR signal reappears in a rather narrow
low-$T$ range. In Fig \ref{fig-3}(b) we show that the intensity is
completely recovered between 1.7~K and 1.2~K at 4.5~T, while at
lower fields it is not fully recovered yet at 1.2~K. This points to
a rapid temperature variation of the threshold field, which is a
strong indication of a field-induced transition in NGS.

On the theoretical side, considerable attention has been recently
devoted to the magnetic-field effects on the ground state of the
Ising-like KAF, either due to easy-axis single-ion anisotropy
(SIKAF) \cite{Xu, Sen} or anisotropic exchange (AEKAF)
\cite{Xu,Moessner,Cabra}. Since the CF symmetry on the
Nd$^{3+}$ site is low in NGS \cite{Bordet}, one expects that the
lowest Kramer's doublet essentially describes the low-$T$ physics. A
priori, one can not discard single-ion nor exchange anisotropy. If
the exchange $J=3|\theta_{CW}|/4S(S+1)\approx1.5$~K is much lower
than the CF splitting between the ground and excited states the
SIKAF picture will apply, otherwise the exchange anisotropy may be
important. For this latter AEKAF case, an "exotic ferromagnetic"
phase, showing finite magnetization but no conventional order, was
predicted at $H=0$ with a transition temperature $T_c \lesssim 0.08
J_{z} S/A$, which depends on the asymmetry level $A=J_z/J_{xy}$
\cite{Kuroda}. To comply with the absence of transition above 60~mK
in NGS, this model would yield a strong easy axis scenario with
$A\gtrsim10$ which does not fit well with the moderate magnetocrystalline
anisotropy observed at low $T$ \cite{Bordet}. NGS is thus likely best described in
the SIKAF model with finite single ion anisotropy $D$. The
susceptibility crossing at 33~K suggests that $D$ is of the order of 10~K. Our NMR results at 1.2~K together with magnetization data at
1.6~K~\cite{Bordet} point to a magnetization plateau with 1.6~$\mu_B$
per Nd$^{3+}$. It is worth noting that the expectation value of
the total angular momentum operator within the ground Kramer's
doublet is likely reduced with respect to the full free-ion value (3.27~$\mu_B$). Therefore, the magnetization fraction $M/M_f$ on the observed plateau
is not accurately known, but $M/M_f \gtrsim 1/2$. In striking
contrast, all anisotropy models on the kagom\'e lattice predict a
1/3-magnetization plateau \cite{Sen,Moessner,
Cabra}. This plateau should be stable up to a field of ca. 10~T ($H\sim J S$) \cite{Sen}, in contrast with the experiment.

In conclusion, our results point to the persistence of the fluctuating
disordered state in NGS down to 60~mK in zero field. The rather slow fluctuations, $\nu_e \approx 4$~GHz, witness a collective paramagnetic state below 10~K, which is a new relevant energy scale in the system. NGS and
ZnCu$_3$(OH)$_6$Cl$_2$ thus enable an experimental verification of
theoretical predictions of a spin-liquid ground state for quite
different Hamiltonians. The fluctuations in NGS are considerably
suppressed by the applied field and signify an unpredicted
field-induced phase transition above 0.5~T at 60~mK. Much like in
rare-earth based pyrochlores, our results speak in favor of
competing single ion anisotropy and exchange interaction. Such a
competition may lead to the observed collective paramagnetism and to
its instability against the applied magnetic field.

We acknowledge discussions with G. Misguich and V. Simonet. We thank A. Amato and C. Baines for technical support at PSI. The work was partially supported by the EC Marie Curie Fellowship (MEIF-CT-2006-041243), by the EC FP 6 program (RII3-CT-2003-505925) and by the ANR project "OxyFonda" (NT05-4\_41913).

\appendix


\begin{thebibliography}{000}
\bibitem{foot3} We performed a dipolar-field calculation by taking account of the Nd$^{3+}$ moments within a fictitious Lorentzian sphere around a particular site, so that the dipolar field would converge. This field was then combined with a continuum theory by adding a macroscopic Lorentzian field.\label{foot3}
\bibitem{foot4} Since $\lambda,1/T_1 \propto (\gamma H_{l})^2$, the
amplification of the muon relaxation by a factor of 220 with respect
to the nuclear relaxation is in accordance with $\gamma_\mu /^{71}\gamma
= 10.44$ and a somewhat larger local magnetic field $H_{l}$ at muon
(oxygen) sites than at the Ga(3) site.\label{foot4}
\end{thebibliography}
\end{document}